\documentclass{article}

\usepackage{microtype}
\usepackage{graphicx}
\usepackage{subfigure}
\usepackage{booktabs} 

\usepackage{hyperref}

\usepackage[accepted]{icml2021}

\icmltitlerunning{PromptCrafter: Crafting Text-to-Image Prompt through Mixed-Initiative Dialogue with LLM}

\begin{document}

\twocolumn[
\icmltitle{PromptCrafter: Crafting Text-to-Image Prompt through Mixed-Initiative Dialogue with LLM}



\icmlsetsymbol{equal}{*}

\begin{icmlauthorlist}
\icmlauthor{Seungho Baek}{kaist-sc}
\icmlauthor{Hyerin Im}{kaist-id}
\icmlauthor{Jiseung Ryu}{kaist-sc}
\icmlauthor{Juhyeong Park}{kaist-id}
\icmlauthor{Takyeon Lee}{kaist-id}
\end{icmlauthorlist}

\icmlaffiliation{kaist-sc}{School of Computing, KAIST, Daejeon, Korea}
\icmlaffiliation{kaist-id}{Department of Industrial Design, KAIST, Daejeon, Korea}

\icmlcorrespondingauthor{Seungho Baek}{sk\_and\_mc@kaist.ac.kr}

\icmlkeywords{mixed-initiative, question-answer, text-to-image generation, prompt engineering, large language model}

\vskip 0.3in
]



\printAffiliationsAndNotice{}  

\begin{abstract}
Text-to-image generation model is able to generate images across a diverse range of subjects and styles based on a single prompt.
Recent works have proposed a variety of interaction methods that help users understand the capabilities of models and utilize them.
However, how to support users to efficiently explore the model's capability and to create effective prompts are still open-ended research questions.
In this paper, we present \textsl{PromptCrafter}, a novel mixed-initiative system that allows step-by-step crafting of text-to-image prompt. Through the iterative process, users can efficiently explore the model's capability, and clarify their intent.
\textsl{PromptCrafter} also supports users to refine prompts by answering various responses to clarifying questions generated by a Large Language Model. Lastly, users can revert to a desired step by reviewing the work history.
In this workshop paper, we discuss the design process of \textsl{PromptCrafter} and our plans for follow-up studies.
\end{abstract}

\section{Introduction}
\label{introduction}

\begin{figure}[h!]
  \centering
  \includegraphics[width=0.45\textwidth]{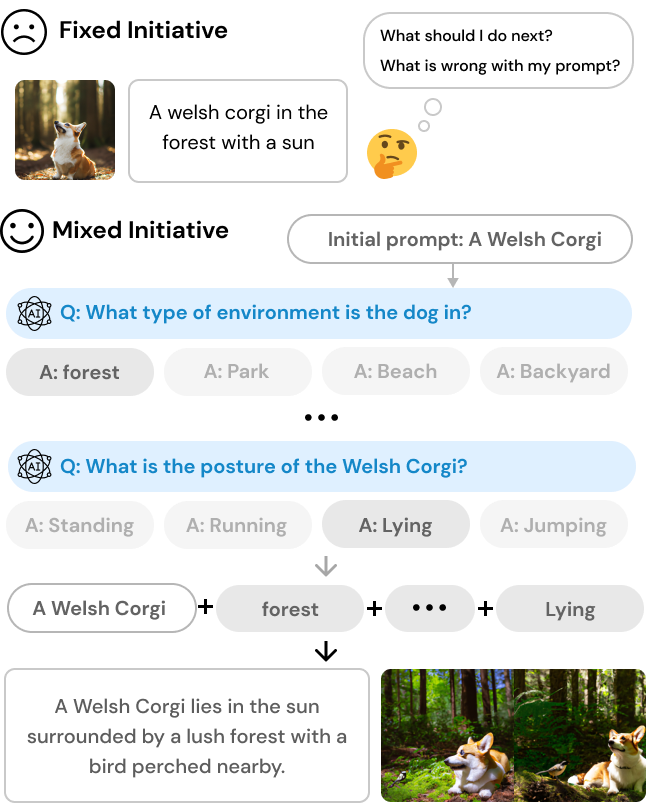}
  \caption{To generate the desired image using a text-to-image generation model, it is necessary to modify the completed prompt by adding or removing content. In the existing fixed-initiative approach, it is difficult to determine which keywords to add or remove from a long prompt. In this work, we decompose prompt writing into category units and generate images using a mixed-initiative approach that answers AI questions.}
  \label{fig:teaser}
\end{figure}

The advancement in computer vision technology and the availability of large amounts of training data have led to the emergence of Large-scale Text-to-image Generation Model (LTGM) such as DALL-E series\cite{dalle_ramesh_2021, dalle2}, Midjourney\cite{midjourney}, and Stable Diffusion\cite{stable-diffusion}.
LTGM has received significant attention for its ability to generate feasible images from a single text prompt, and has begun to be used in design workflow \cite{3dalle}.
In particular, due to its use of free-form text to leverage almost infinite generation capability, it possesses an almost limitless potential for creation.
However, since it cannot be guaranteed which prompt will produce a quality outcome, the user's design process for generating an image often involves a trial and error process.
In order to address this issue, recent research has explored structured prompts that produce high-quality images \cite{design-guideline} and proposed a structured exploration system to support understanding generative AI capabilities \cite{opal}. 

Prompt engineering, appropriately modifying prompts in order to obtain the specific results that user desire, is necessary in addition to creating good prompts \cite{prompt-engineering}.
Many real-world tasks involve an iterative process of adding, modifying, and subtracting parts, rather than just a single model run, because it is difficult to have a concrete idea and make an accurate request that reflects it. 
The structured approach allows for the creation of high-quality images easily, but it can be challenging to incorporate information beyond the predetermined structure when creating an image.
Furthermore, identifying problematic keywords and modifying them can be challenging especially with a lengthy prompt.

In this paper, we propose a novel mixed-initiative interaction approach with Large Language Model (LLM), in which we decompose a prompt into smaller steps through Question-Answer (QA), rather than modifying a complete prompt in a fixed-initiative manner, as shown in Fig.\ref{fig:teaser}.
This approach allows users to refine prompts by answering various responses to clarifying questions generated by a Large Language Model.
We design and develop \textsl{PromptCrafter}, a system that provides step-by-step QA and visualizes the QA histories, allowing users to explore the capabilities of a model while interacting with it and to understand and modify the prompt engineering process effectively.
Through \textsl{PromptCrafter}, users can craft text-to-image prompt by exploring a variety of results through QA and can easily revise and make changes to their workflow as desired.
In this workshop paper, we discuss the design process of \textsl{PromptCrafter} and our plans to study our research questions.

\begin{figure*}[ht]
\begin{center}
\centerline{\includegraphics[width=0.95\linewidth]{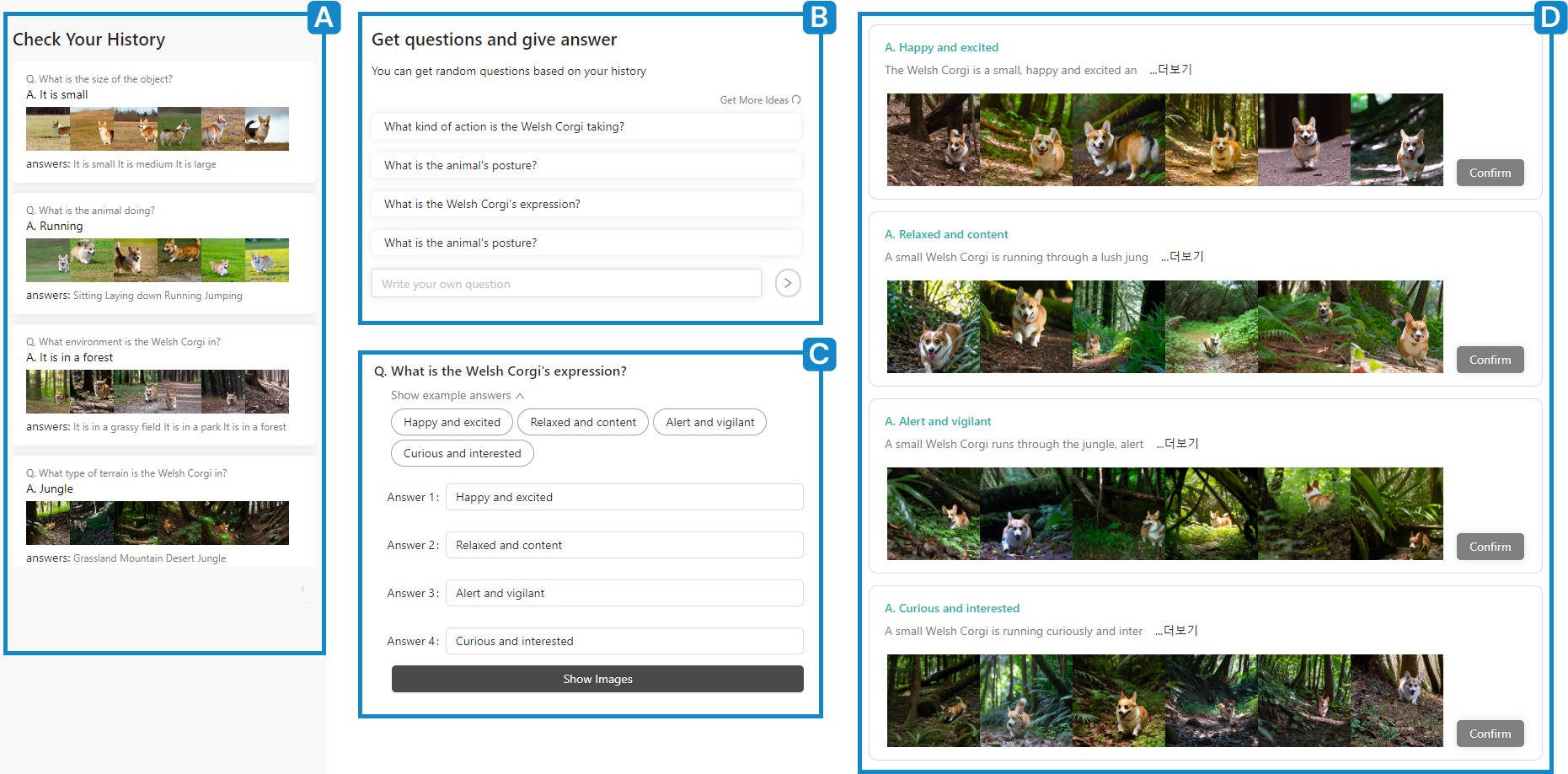}}
\caption{\textsl{PromptCrafter} is  a novel mixed-initiative system that allows step-by-step crafting of text-to-image prompt through questions and answers.In the \textsl{PromptCrafter}, user first input the initial prompt, such as a welsh corgi, and \textsl{PromptCrafter} then provides clarifying questions such as 'What is the posture of the Welsh Corgi?' (B) and \textsl{PromptCrafter} suggests four sample answers for selected question to provide inspiration, and user can either use them or enter their own responses (C). \textsl{PromptCrafter} then generates images using GPT-3 based on the initial prompt and the question-answer histories (D). When the user confirms the desired result, the step is completed and the QA record is saved in history (A). Based on this, a new question is presented and a new step begins. \textsl{PromptCrafter} then generates new images based on the image concept and the user's question-answer histories. Throughout the process, the user can explore the image generation model and clarify ideas to generate the desired image.}
\label{fig:system}
\end{center}
\end{figure*}
\section{Design Process}
\label{design}

In order to understand the difficulties encountered in generating images using LTGM, we conducted a formative study. To identify common difficulties regardless of expertise in image creation, we recruited a total of 18 participants, with 6 participants each in three skill levels (novices, hobbyists, and experts). We asked each participant to use DALL-E2 to create an image that represents ``sustainable future lifestyle''. Based on recorded screen videos and post interviews, we collected common difficulties that LTGM users would experience while generating images, and defined three design problems as below. 

\textbf{P1: Hard to identify a part of a prompt that causes undesired results, and to make necessary modifications.} 
While users could easily create images by crafting prompts containing various information (e.g., A welsh corgi in the forest with a sun), it was much harder to make adjustments when they got unexpected or undesired results. Most existing LTGM services do not provide relevant information for modifying or removing keywords within the prompt.
Thus, users tend to rely on intuition to identify problematic parts and to fix them. Moreover, when certain keywords are deleted, the entire sentence also need to be restructured.

\textbf{P2: Compromise on incomplete outcomes rather than adding content that cannot guarantee better results and have difficulty in prompt engineering.} 
No participants could get desired images in a single trial. Instead they went through multiple rounds of minor tweaks such as adding, removing, and replacing keywords. Unfortunately many participants failed to get satisfying results, and chose to end up with compromised outcomes, thinking that major changes that may cause even more problems.

\textbf{P3: Difficult to utilize LTGM's capability in the workflow, as only certain prompts can be used due to the issue with P2.}
The users tend to use some prompts that have produced good results instead of generating images by changing prompts in a diverse based on the results from P2.
As a result, it becomes difficult to fully understand the wide-ranging abilities of LTGM and only limited functionalities are utilized.

Based on the defined problems, we set three design goals for a LTGM based image-generation system.

\begin{itemize}
    \item G1. Decomposing prompt completion into smaller steps that deal with single idea to make problem identification and correction easier.
    \item G2. Providing examples and multiple results to be compared simultaneously in order to explore various image generation directions.
    \item G3. Visualizing the workflow enables the easy understanding of the LTGM working process and supports for the generation of various results.
\end{itemize}

\section{PromptCrafter}
\label{system_design}

In this section, we introduce the rationale for selecting Question-Answer (QA) as our design process outcome, and the interface of \textsl{PromptCrafter}.

\subsection{Question and Answer}
To achieve the design goals, we decompose the image generation process into multi-steps (G1) and utilize QA at each step to effectively explore the models and clarify user's ideas (G2).
We used QA because it has been researched that it allows for retrieving exact answers for users\cite{qa-review} and helps users understand the system\cite{sara, mathbot}, while a mixed initiative user interface enables efficient collaboration between users and intelligent agents\cite{mixed-initiative}.
Furthermore, based on the user's workflow history, the GPT-3\cite{gpt3}, which is one of the LLM, provides clarifying questions and sample answers in order to help user explores LTGM's capabilities easily.
Finally, the LLM generates prompts for image generation by utilizing the QA histories and also the QA structured prompt generates high-quality writing \cite{helpmethink, chainofthought}.
The user can focus on each step without having to worry about writing prompts.
Therefore, we employ a mixed-initiative approach utilizing QA as a suitable solution to address our design goal.

\subsection{Interface}

\subsubsection{Question-Answer Panel}
\textsl{PromptCrafter} provides the question panel (Fig.\ref{fig:system}B) that proposes four (or even more on request) LLM-generated clarifying questions. By selecting the most interesting question, users decide what information to be added at the current step - i.e., how to expand / clarify their intent. For instance, ``what type of environment is the dog in?'' is generated by LLM in consideration of the initial prompt and QA histories. User can click on 'Get More Ideas' to receive different questions or write their desired question in the input field. When a question is selected, LLM proposes four sample answers in the answer panel (Fig.\ref{fig:system}C). Users can choose or manually type up to four answers. Lastly, users click the ``Show Images'' button to generate six images for each answer they chose.  



\subsubsection{Confirmation Panel}
The confirmation panel (Fig.\ref{fig:system}D) shows six images for each answer based on previous QA histories and current responses. Users can easily compare the generated images, and press ``Confirm'' button of the most desired set of images. The answer of the chosen image set is automatically applied to the prompt (G2).
\textsl{PromptCrafter} also supports the creation of prompts for image generation through the use of LLM in this process.
Based on the QA, LLM generates prompts for generating images and LTGM generates images using the resulting prompts.
The user can review the prompts and images generated for each answer and can additionally change some answers to other responses while searching for desired images.
When the user finds the desired outcome, user can click the 'confirm' button to save the work up to this point and proceed to the next step.

\subsubsection{History Panel}
The history panel (Fig.\ref{fig:system}A) shows the list of previous steps, and allows users to return to any step, in order to create images of different topics (G3).
Each step in the history panel contains the generated images for the selected question and the confirmed answers, along with remaining answers.
Users can review those images that they previously created for the step, but did not confirm by clicking them.
And if desired, user can also revert back to that step to explore other answers, compare them, and create a different image than the initial work.

\begin{figure}
    \begin{center}
        \centerline{\includegraphics[width=0.93\linewidth]{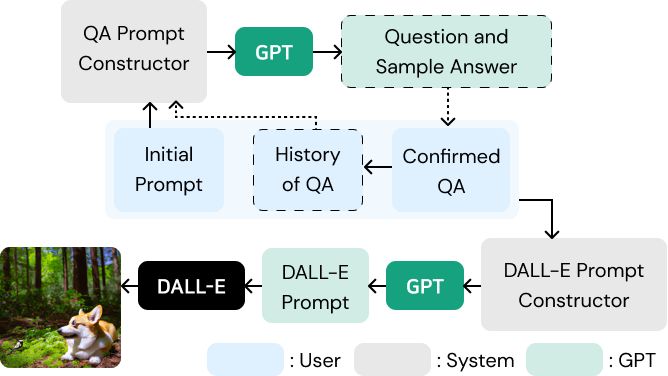}}
        \caption{The overview of \textsl{PromptCrafter} architecture. The user's confirmed QA is saved in the history when the step is completed. And when the next step begins, the system constructs a prompt based on the history and initial prompt in order to generate a question and sample answer. After the user selects the questions and answers, a DALL-E prompt is constructed based on the previous user's data, and images are generated through the DALL-E prompt generated by the GPT.}
        \label{fig:architecture}
    \end{center}
\end{figure}

\subsection{Implementation}
\textsl{PromptCrafter} is a web application that was developed using HTML/CSS/JS in the React framework for its interface, and communicates with a back-end server that operates an AI model through API-based data exchange.
The back-end server was developed using the express framework in Node.js. After receiving information inputted or selected by the user through the interface, we utilize OpenAI's GPT and DALL-E API\footnote{https://openai.com/blog/openai-api} to generate results, which are then returned to the interface through an API.
As shown in Fig.\ref{fig:architecture}, the prompt constructor utilizes the user's initial prompt and QA to construct a prompt, allowing GPT to generate clarifying questions, sample answers, and DALL-E prompts.
When the step is completed, the user's QA is saved in the history and GPT asks questions for more specific information based on it.
\section{Future Work}
\label{future_work}

We plan to evaluate a user study to verify if the use of LTGM in the \textsl{PromptCrafter} process is effective.
The current system has been developed and is ready for experimentation. Additional development may be necessary if supplementary features, such as logging user-specific behavior during the study design process, are required.
The research goals for the user study are: 1) Does \textsl{PromptCrafter} effectively support the LTGM prompt engineering process?, 2) Does \textsl{PromptCrafter} support the process of clarifying ideas?, and 3) Does \textsl{PromptCrafter} assists in understanding and utilizing LTGM?
To achieve this, we will analyze the results by conducting the same tasks using both the current text-to-image generation system and the \textsl{PromptCrafter} in a between-subjects design.


\bibliography{main}
\bibliographystyle{icml2021}

\end{document}